%% file: new_main.tex
\documentclass[9pt,conference]{IEEEtran} 
\input{packages}

\title{Compressive Electron Backscatter Diffraction Imaging}

\author[1]{Zoë Broad}
\author[2]{Alex W. Robinson}
\author[3]{Jack Wells}
\author[2]{Daniel Nicholls}
\author[1,4]{Amirafshar Moshtaghpour}
\author[4,5]{\\Angus I. Kirkland} 
\author[1,2]{Nigel D. Browning}

\affil[1]{Department of Mechanical, Materials and Aerospace Engineering, University of Liverpool, UK.}
\affil[2]{SenseAI Innovations Ltd., Liverpool, UK.}
\affil[3]{Distributed Algorithms Centre for Doctoral Training, University of Liverpool, Liverpool, UK.}
\affil[4]{Correlated Imaging Group, Rosalind Franklin Institute, Harwell Science and Innovation Campus, Didcot, UK.}
\affil[5]{Department of Materials, University of Oxford, Oxford, UK}

\begin{document}     

\maketitle          

\begin{abstract}   
    Electron backscatter diffraction (EBSD) has developed over the last few decades into a valuable crystallographic characterisation method for a wide range of sample types.
    Despite these advances, issues such as the complexity of sample preparation, relatively slow acquisition, and damage in beam-sensitive samples, still limit the quantity and quality of interpretable data that can be obtained.
    To mitigate these issues, here we propose a method based on the subsampling of probe positions and subsequent reconstruction of an incomplete dataset. 
    The missing probe locations (or pixels in the image) are recovered via an inpainting process using a dictionary-learning based method called beta-process factor analysis (BPFA).
    To investigate the robustness of both our inpainting method and Hough-based indexing, we simulate subsampled and noisy EBSD datasets from a real fully sampled Ni-superalloy dataset for different subsampling ratios of probe positions using both Gaussian and Poisson noise models.
    We find that zero solution pixel detection (inpainting un-indexed pixels) enables higher quality reconstructions to be obtained.
    Numerical tests confirm high quality reconstruction of band contrast and inverse pole figure maps from only 10\% of the probe positions, with the potential to reduce this to 5\% if only inverse pole figure maps are needed. 
    These results show the potential application of this method in EBSD, allowing for faster analysis and extending the use of this technique to beam sensitive materials.
\end{abstract}

\section{Introduction}

\begin{figure}
	\centering
	\includegraphics[width=1\columnwidth]{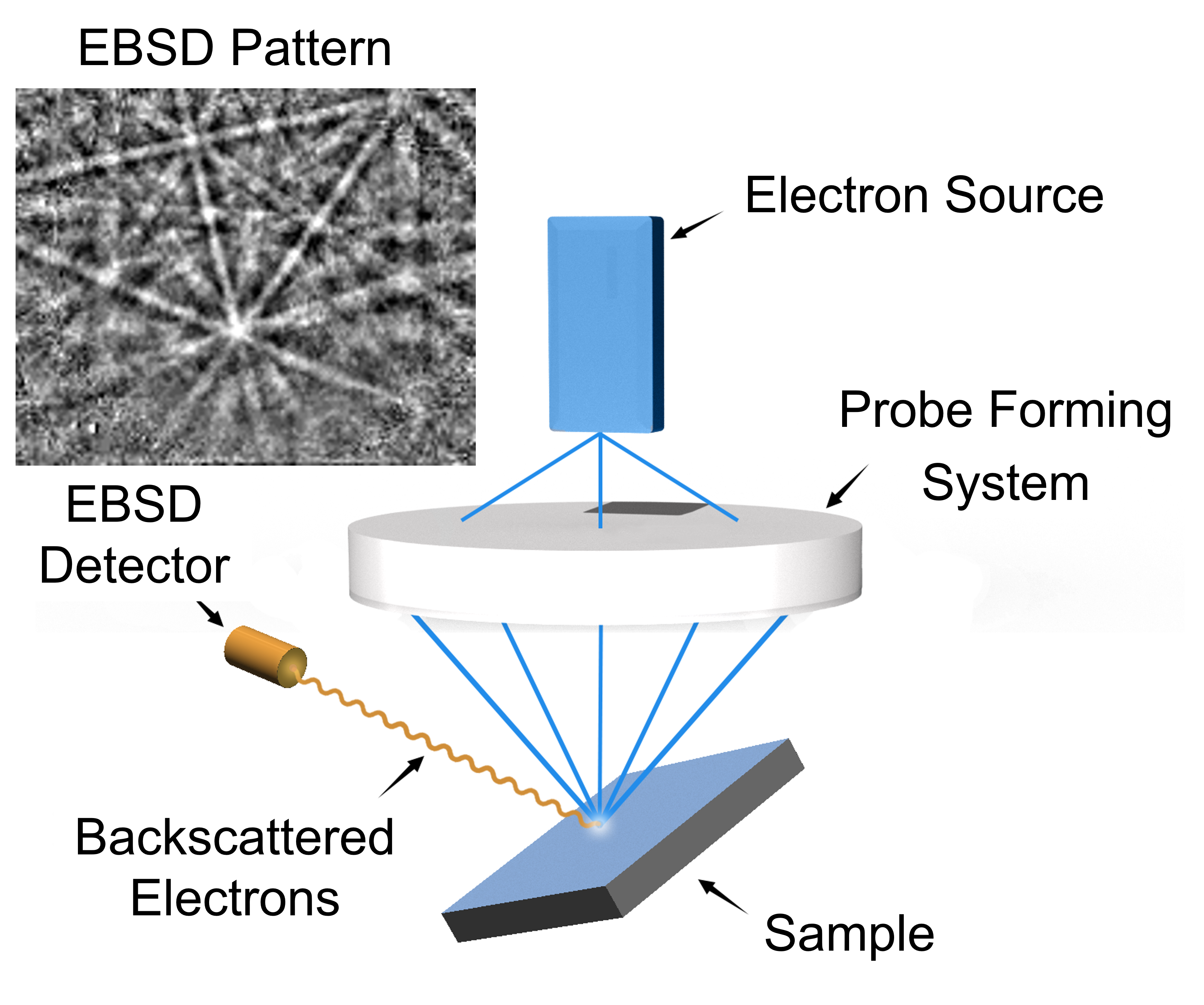}
	\caption{\textbf{Operating principles of EBSD imaging.} A convergent electron beam is raster scanned across the sample. Backscattered electrons form a pair of cones which intersect the phosphor screen, allowing the EBSD pattern to be read by the detector.}
	\label{fig:ebsd-schematic}
\end{figure}

Electron backscatter diffraction (EBSD) is a scanning electron microscopy (SEM) technique, providing important crystallographic information about the sample such as crystal orientation and grain size~\cite{EBSDinMatSci}. 
% \Rc{odd choices of reference - recommend Schwartz et al}. 
Figure~\ref{fig:ebsd-schematic} shows the typical instrument geometry, where an EBSD pattern (EBSP) is formed from an electron beam incident on a crystal plane of a highly tilted sample~\cite{EBSDinMatSci}. 

Conventionally, an EBSD dataset is acquired by scanning an electron beam across the surface of a sample in a raster scan, acquiring an EBSP at every position in the scan, \ie at each probe location. The measured EBSPs form a 4-dimensional (4-D) dataset~\cite{EBSDinMatSci}.
% \Rc{clarify why we're calling it 4-D}

\begin{figure}
	\centering
	\includegraphics[width=1\columnwidth]{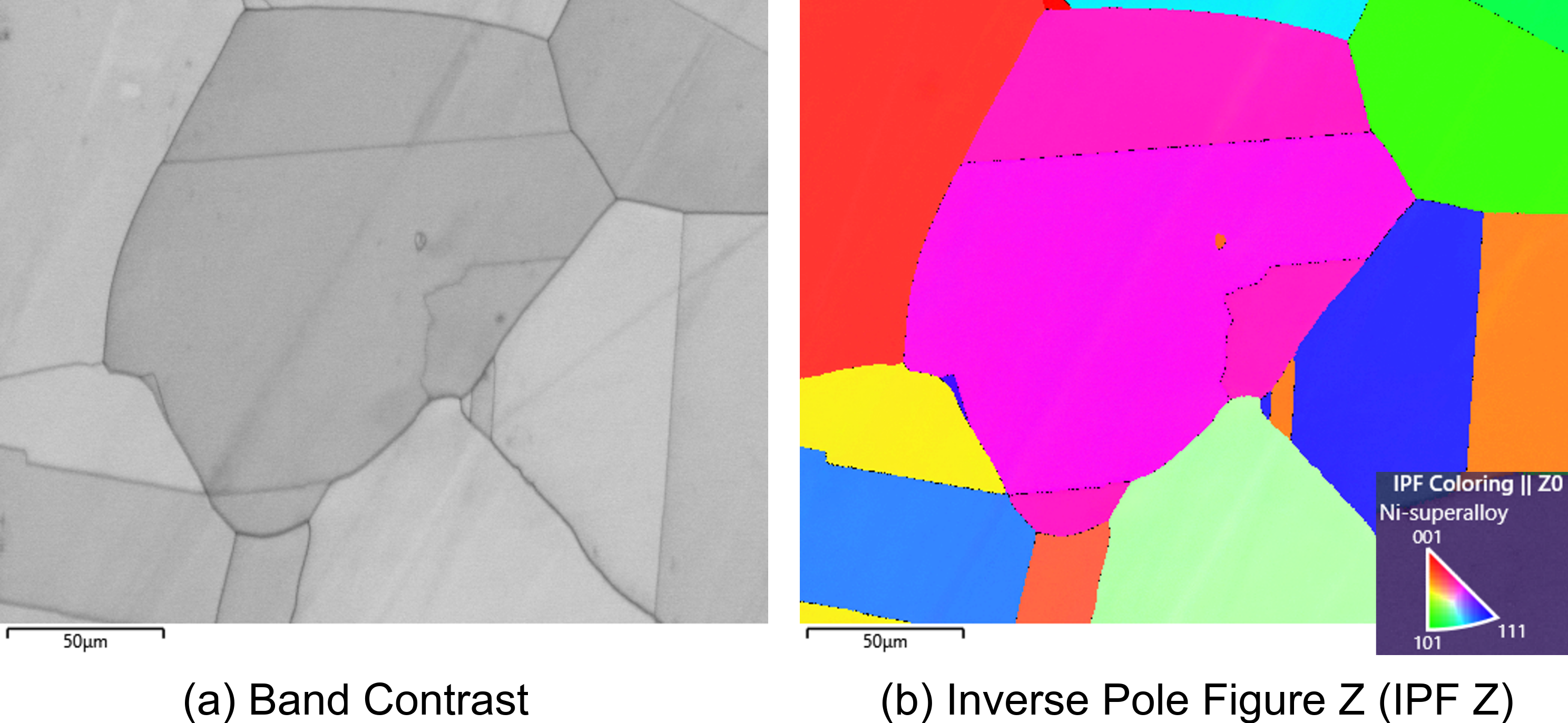}
	\caption{\textbf{Example EBSD Maps of Ni-Superalloy.} EBSD maps provide crystallographic information about the sample such as (a) the locations of grain boundaries or (b) crystal orientation.}
	\label{fig:mapex}
\end{figure}

EBSPs contain information relating to crystal structure, and this information is extracted by indexing -- \ie the process of identifying the crystal phase and orientation from each EBSP. Multiple methods exist to allow this process to be performed automatically ~\cite{hough1962method, lassen1994automated, chen2015dictionary, lenthe2019spherical, rowenhorst2024fast}. Each EBSP corresponds to a single probe position in the sample and therefore to a single pixel in the indexed map. Each pixel in this map then represents some information gained from each EBSP. Band contrast and inverse pole figure (IPF) maps are two common types of EBSD maps, which are chosen as the focus of study in this work (Figure ~\ref{fig:mapex}).

Band contrast maps are a measure of EBSP quality based on the contrast between the intensity of the Kikuchi bands compared to the intensity of the background in an EBSP. At grain boundaries and defects the contrast is lower due to the superimposition of EBSPs from all grain boundaries in the interaction volume~\cite{wright2006ebsd}. This gives rise to a low indexing value in the band contrast map, producing dark pixels at the grain boundaries. This provides additional crystallographic information to be obtained compared to conventional secondary/backscattered electron imaging due to this mechanism~\cite{SEMGoldstein}. Other quality maps, \eg image quality and band slope maps provide similar information, and are expected to give equivalent results to those demonstrated here since they have the same image properties.
% \Rc{this is from a reviewer comment} 
%\ZBc{Amir comments to here 6/6/24 14:25}
Orientation maps such as an IPF map are a measure of the relative crystal orientation at each probe position. For example IPF Z map, also called a normal direction map, in Figure~\ref{fig:mapex}. In an IPF map the colour of each point corresponds to the crystallographic direction with regards to a cartesian axis, \ie the Z axis in the case of the IPF Z map.

% , Z refers to the axis from which crystal orientation is measured; the legend shows how the colours correspond to the relative orientation measurement \eg a blue grain corresponds to a [111] crystal orientation, with respect to the crystal frame, whilst a blue/green colour corresponds to [212]. \Rc{IPF definition needs clarification. This type of map is more typically referred to as a normal direction IPF map as it shows the crystal direction aligned with the sample normal.}

Despite significant advances in EBSD imaging, currently its applications remain limited due to, for example, sample preparation. Since a perfectly flat surface is needed for quality measurements, complex samples (such as geological samples or samples with many phases where there may be low signal) and detector speed can limit quality and/or quality of a dataset. The fastest detectors currently on the market can reach speeds of over 6000 patterns per second \cite{EDAXCMOS} but this depends on the sample and quality needed. 
% \Rc{specify speeds}

\subsection{Contributions}

\begin{figure*}[t]
	\centering
	\includegraphics[width=2\columnwidth]{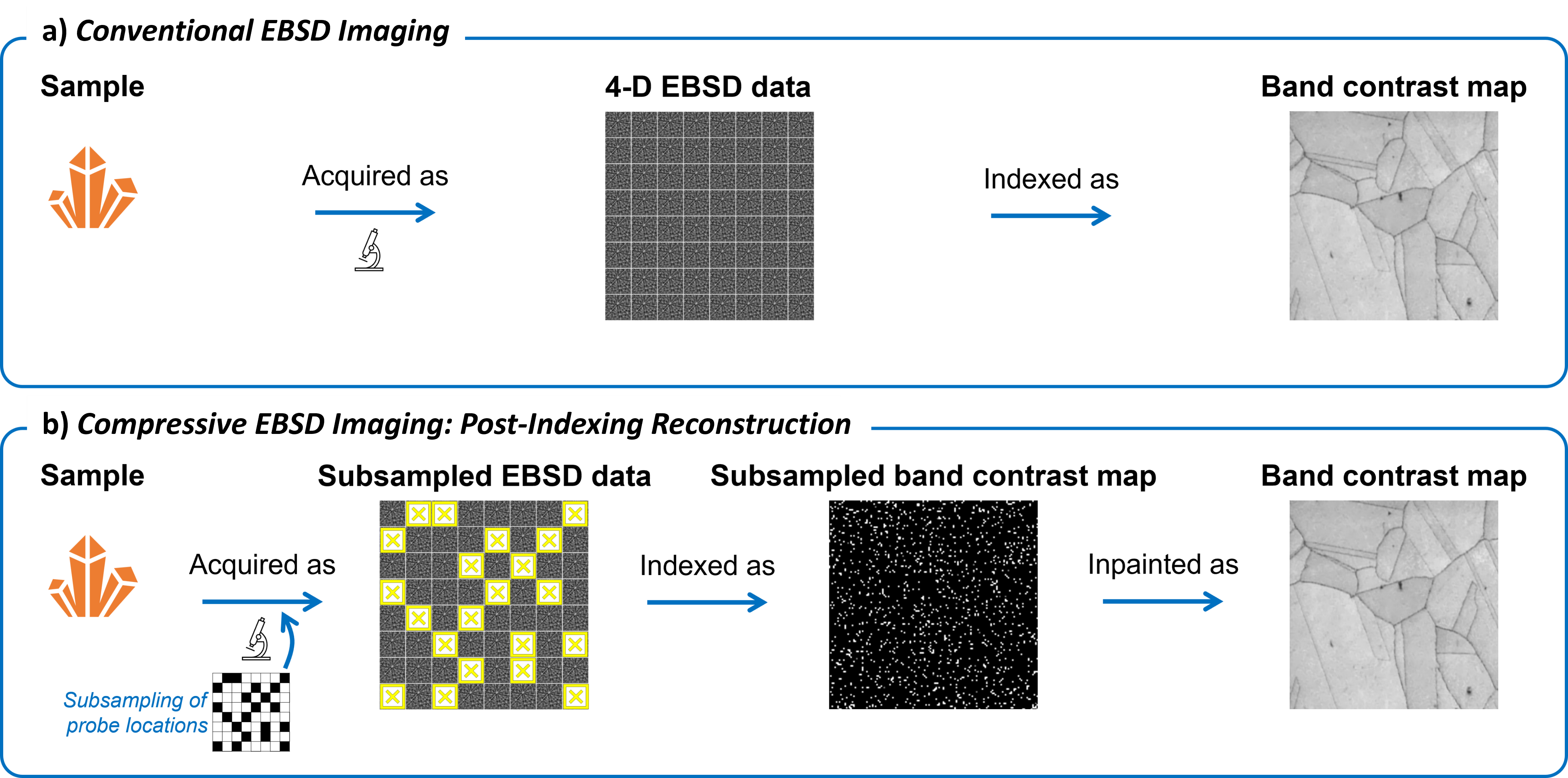}
	\caption{\textbf{Conventional \textit{vs.} compressive EBSD imaging.} Probe locations are subsampled and the EBSD patterns acquired are indexed. The incomplete maps formed can then be inpainted.}
	\label{fig:ebsd_workflow}
\end{figure*}

In this paper, we introduce an EBSD acquisition approach based on subsampling probe positions, which results in an incomplete 4-D EBSD dataset. Subsampling allows significantly faster data acquisition without any significant advances for any detector technology. Indexing such data, however, yields band contrast and IPF maps with missing values associated with unsampled probe positions. Our approach, as illustrated in Figure~\ref{fig:ebsd_workflow} (b), is to first index the subsampled EBSD dataset and then to inpaint the resulting EBSD maps using beta-process factor analysis (BPFA) ~\cite{sertoglu2015scalable, paisley2009nonparametric}, \ie a joint blind dictionary learning and sparse coding image inpainting method.

Here we consider only analyses where maps are produced to visualise sample texture, \eg where only a pattern quality or misorientation map is needed, rather than for the analysis of deformations where the maps produced require knowledge of neighbouring EBSPs. Additionally, it should be noted that the method applied here focusses on improving the speed of EBSD acquisition, since regardless of detector speed subsampling improves acquisition speed by acquiring fewer probe locations. It is anticipated that this could be most useful for identifying areas of interest, large area mapping, or for 3-D EBSD where large quantities of data are being acquired. The latter also has the potential for targeted sampling to be applied, where the sampled datapoints are chosen based on the previous map.

Robustness to subsampling is first investigated here using a real dataset acquired with high fluence. Robustness to noise is evaluated considering both a Gaussian and Poisson noise model; first without subsampling, testing the efficacy of the indexing procedure and then with subsampling, testing the efficacy of the inpainting method.

\subsection{Related works}

\subsubsection{Applications of compressive sensing in electron microscopy}
Compressive sensing (CS) has seen successful applications in several modalities of electron microscopy for, \eg low dose or fast acquisition. The theory of CS consists of a set of conditions for recovering a signal from few direct or indirect measurements compared to the intrinsic dimension of that signal~\cite{candes2006robust, donoho2006compressed}. CS often relies on the fact that most data can be sparsely represented in an appropriate sparsity basis~\cite{mallat1999wavelet}, or dictionary.

CS theory has been applied to multiple modes in electron microscopy. Scanning transmission electron microscopy (STEM) in particular has seen developments in this area, allowing for low dose acquisition to be realised ~\cite{saghi2015reduced, kovarik2016implementing, mucke2016practical,nicholls2022compressive, nicholls2022sub, browning2023advantages} 

CS has enabled fast and low-dose 4-D STEM systems~\cite{robinson2023silico,robinson2023simultaneous,moshtaghpour2023exploring}. It has been demonstrated the application of CS allows sampling rates below the Nyquist criteria to be realised meaning that faster acquisition and lower beam dose can be implemented \cite{robinson2023developing}. STEM simulations have also benefited through the use of CS, allowing close to real time calculation of STEM images~\cite{robinson2023towards,robinson2022sim}. FIB-SEM is another method that has seen CS applications, enabling a lower electron dose acquisition and reducing the long acquisition times ~\cite{nicholls2023targeted, nicholls2023potential}.

\subsubsection{Alternative Scanning Strategies for EBSD Imaging}

%\textit{Alternative Scanning Strategies:}
The method described here is not the first alternative EBSD scanning strategy to be investigated for EBSD.
% Alternative scanning strategies, enabling smaller datasets to be acquired has been previously investigated in EBSD. 
A method of smart sampling, enabling the targeted sampling of grain boundaries was proposed by Yang et al \cite{yang1999adaptive}. Another alternative strategy proposed by Tong et al \cite{tong2019rapid} used the forward scattered electron (FSE) image acquired to identify grains, enabling single orientation measurements of each grain to be analysed. Closest to the method described in this paper is a method of determining grain size distribution proposed by Long \etal~\cite{long2024high}, which uses grid-based, random and quasi-random sampling for data acquisition. In this work the desired output is the grain size distribution, whilst we aim to reconstruct the EBSD maps.

The scanning strategies proposed by Yang et al and Long et al have the desired output of grain boundary information. As in our method, in Tong et al a visualisation of texture is the key output. Our proposed method does not require any prior knowledge of the same \eg through the FSE image, although this is a technique that could be investigated in the future.

\subsubsection{Camera Subsampling:}
In EBSD imaging, a method of camera subsampling has been demonstrated by Wang \etal~\cite{wang2021electron}. This method means that fewer pixels on the detector are acquired, demonstrating the significant speed benefits of subsampling. Our method differs by applying probe subsampling, which has the potential to be implemented in conjunction with camera subsampling, which would enable even faster acquisition. 

%This work differs from this in two aspects. First, probe subsampling is considered, as opposed to only camera subsampling, wherein incomplete EBSPs are acquired. Secondly, BPFA, a joint dictionary-learning and sparse coding algorithm, is used to recover the missing data, rather than an interpolation method. Additionally, the application of probe subsampling does not rule out camera subsampling, with the application of both methods being a possibility that would enable an even faster and efficient analysis. 

\subsubsection{Cleaning strategies:}
Many methods are available for EBSD imaging, allowing samples with high levels of noise to be more accurately indexed, \eg by using information from neighbouring pixels, as in methods like grain dilation and neighbouring pattern averaging and reindexing (NPAR) or through alternative post-processing of the full dataset, such as non-local means pattern averaging and reindexing (NLPAR) or multivariate statistical analysis \cite{wright2006random, brewick2019nlpar, brewer2008multivariate, wilkinson2019applications}. Although, as addressed in Brewer et al \cite{brewer2010risks}, there are risks of overprocessing the datasets which can result in misindexing or the formation of artefacts.

In this work we detect zero solution pixels, which are pixels where the EBSP has not been indexed by the indexing algorithm, and consider them to be an unsampled data-point, which allows them to be inpainted in the same manner as the intentionally unsampled locations. Our method is most similar to that of grain dilation \cite{wright2006random} since it is based on the information of the neighbouring points. They key difference being that we clean the data as it is being inpainted. 
%This method prevents the zero solution pixels from being considered as part of the image \eg as a measured orientation. Although this has the benefit of speed and from learning from the full map it does not have the added benefit of considering the EBSD patterns themselves in this method, such as in NLPAR and in the multivariate statistical analysis methods. For datasets which do not have problems with severe noise or misindexing this is unlikely to be a problem, however it does mean that there are datasets which may not be well suited for this method of analysis, particularly where speed and beam damage are not considered to be issues. \ZBc{Amir had corrections for this section but since I have absolutely no idea what I'm trying to say I've not done them}

\section{Conventional EBSD imaging and analysis}\label{sec:conventional-ebsd}
In this section for convenience we describe the conventional EBSD acquisition and analysis methods.

% --------------- Subsection 2.1 ------------
\subsection{Acquisition model}\label{subsec:conventional-ebsd-acquisition}
A 4-D EBSD dataset is a collection of 2-D EBSD patterns for every probe position. Consider a probe scanning $H_{\rm p}\times W_{\rm p}$ positions on a regular 2-D grid with $N_{\rm p} \coloneqq H_{\rm p} W_{\rm p}$ total probe positions. For each position of the probe, indexed by $l \in \{1,\cdots, N_{\rm p}\}$, the detector records a 2-D EBSD pattern over $H_{\rm d} \times W_{\rm d}$ pixels. We represent that EBSD pattern in its vectorised format $\bs y_l \in \bb R^{N_{\rm d}}$, where $N_{\rm d} \coloneqq H_{\rm d} W_{\rm d}$ is the total number of detector pixels.

Let $\bs x \in \bb R^{N_{\rm p}}$ be a discretised and vectorised representation of an EBSD map, \eg band contrast map or IPF map, where every pixel corresponds to one probe position. Without the need to specify an actual forward model ~\cite{chen2015dictionary,singh2017application}, we assume that every EBSD pattern 
$\bs y_l$ is related to its corresponding pixel value in $\bs x$ through a sensing operator $\cl A: \bb R \mapsto \bb R^{N_{\rm d}}$. Mathematically, for every probe position $l \in \{1,\cdots,N_{\rm p}\}$,
\begin{equation}\label{eq:forward-model}
    \bs y_l = \cl A(x_l) \in \bb R^{N_{\rm d}},
\end{equation}
where an additive or non-additive noise model is included in $\cl A$.
% --------------- Subsection 2.2 ------------
\subsection{Analysis of complete dataset}\label{subsec:conventional-ebsd-analysis}
Once an EBSD dataset has been acquired, additional processing, known as indexing, is necessary to extract useful crystallographic information from the EBSPs. Discussed here are different methods of indexing and their benefits.

Indexing refers to the identification of the crystal phase and orientation in the acquired EBSP.
In Hough-based indexing, EBSPs are indexed through a Hough transform, converting the bright bands of the EBSP into bright spots in Hough space ~\cite{wright1992automatic, lassen1994automated}. This method is fast and allows for online indexing. However, the greatest disadvantages of the Hough transform are its sensitivity to noise and lower accuracy compared to alternative such as dictionary indexing ~\cite{chen2015dictionary}. When noisy EBSPs are obtained, the hit rate (\ie the percentage of EBSPs that have been indexed) of indexing based on Hough transform significantly decreases, resulting in incomplete EBSD maps ~\cite{chen2015dictionary}.

Dynamical simulations of EBSPs can be created based on a simplified model describing the interaction of the electrons in the sample, creating a spherical master EBSP of predicted reflections based on the sample properties ~\cite{EBSDinMatSci}. These simulations offer great benefits through alternative indexing techniques, such as dictionary and spherical indexing.

Dictionary indexing uses the master EBSP to create a set of templates followed by identifying the acquired EBSP orientation based on those templates ~\cite{chen2015dictionary}. Although this method is less sensitive to noise, it is computationally expensive. The sizes of the dictionaries used increases as sample symmetry decreases, which further extends the computational cost and analysis times ~\cite{chen2015dictionary, ram2017error}.

Spherical indexing uses a similar technique of EBSP matching. In this model the EBSP is projected onto a sphere, allowing greater correlation in EBSP matching. This technique has shown promising results for noisy EBSPs, with increased speeds compared to dictionary indexing and greater robustness against noise than Hough transform based methods ~\cite{lenthe2019spherical}. 

% \ZBc{textit{I'm not sure where to put this }Since the focus of this compressive EBSD method is to enable higher speed acquisition, the Hough transform is used for indexing of datasets due to its speed benefits. }

The indexing approaches above treat one EBSP at a time. Therefore, regardless of the indexing method used, an estimation of the true EBSD map, \ie $\hat{\bs x} \approx \bs x$ is formed by indexing every EBSP denoted by an operator $\Delta: \bb R^{N_{\rm d}} \mapsto \bb R$, \ie for every EBSP indexed $l \in \{1,\cdots, N_{\rm p}\}$,
\begin{equation}\label{eq:indexing-model}
    \hat{x}_l = \Delta(\bs y_l) = x_l + n_l,
\end{equation}
where $n_l$ represents indexing error.
% \begin{remark}\label{remark:zero-solution}
    We note here that due to discrepancy sources in the forward operator $\cl A$ in Eq.~\eqref{eq:forward-model}, the indexing process may fail. The pixels for which the indexing fails are referred to as \textit{Zero-Solution Pixels} (ZSPs). The indices of those pixels are collected in a ZSP set $\Omega_{\rm zsp} = \{l\in \{1,\cdots,N_{\rm p}\}: \Delta(\bs y_l) = 0\}$.
% \end{remark}

\section{Proposed Method: Post-indexing Reconstruction of Subsampled EBSD Data}\label{sec:compressive-ebsd}

% Described here are the proposed compressive EBSD acquisition model and recovery method after indexing the incomplete dataset.
% -------------------- Subsectio 3.1 -----------------
\subsection{Acquisition model}\label{subsec:compressive-ebsd-acquisition}
The proposed compressive EBSD system operates by subsampling the probe positions as has been applied in the SEM by use of a scan generator, as shown in ~\cite{nicholls2023potential}. 

Let $\Omega \subset \{1\cdots,N_{\rm p}\}$ be a subset of $|\Omega| = M_{\rm p}$ probe positions. The ordering of the elements in $\Omega$ controls the trajectory of the scanning probe. Therefore, the acquisition model of compressive EBSD imaging is identical to Eq.~\eqref{eq:indexing-model}, but only for probe positions $l \in \Omega$. Since the acquisition time of an EBSD data set is proportional to the number of scanned probe positions, by subsampling $M_{\rm p}$ probe positions the acquisition time will be reduced by a factor of $M_{\rm p}/N_{\rm p}$ compared to the full acquisition. The total electron fluence is also reduced by subsampling probe positions. Our model supports any arbitrary subsampling strategy, such as Uniform Density Sampling (UDS)~\cite{nicholls2023targeted} -- \ie selecting a probe position uniformly at random, linehop~\cite{nicholls2022compressive} -- \ie subsampling at random the locations adjacent to the probe’s default line trajectory, or dynamic sampling ~\cite{godaliyadda2017supervised, kandel2023demonstration}.

% -------------------- Subsection 3.2 -----------------
\subsection{Analysis of an incomplete dataset}\label{subsec:compressive-ebsd-analysis}

Given a subsampled 4-D EBSD data, an EBSD map $\bs z \in \bb R^{N_{\rm p}}$ can be computed using an indexing operator $\Delta$. Similar to conventional EBSD imaging, the indexing process follows Eq.~\eqref{eq:indexing-model} for sampled probe locations. However, the value of an EBSD map for unsampled probe positions reads zero. Mathematically, 
\begin{equation}
    z_l = \begin{cases}
        \hat{x}_l, & {\rm if~} l \in \Omega,\\
        0, & {\rm if~} l \not \in \Omega.
    \end{cases}
\end{equation}
From Eq.~\eqref{eq:indexing-model}, this becomes
\begin{equation}\label{eq:indexing-model-compressive}
    \bs z = \bs P_{\Omega}(\bs x + \bs n), 
\end{equation}
where $\bs n \coloneqq [n_1,\cdots, n_{N_{\rm p}}]^\top$ is a noise vector collecting indexing errors; and $\bs P_\Omega \in \{0,1\}^{N_{\rm p} \times N_{\rm p}}$ is a mask operator of probe positions, such that $(\bs P_\Omega \bs x)_l = x_l$ if $l\in \Omega$ and $(\bs P_\Omega \bs x)_l = 0$, otherwise.

% \begin{remark}\label{remark:zero-solution-subsampling}
%     Recalling Remark~\ref{remark:zero-solution}, 
    We note that ZSPs result in zero values in vector $\bs z$. Depending on the application, those ZSPs can be detected and treated as unsampled pixels, \ie $\Omega \leftarrow \Omega \cup \Omega_{\rm zsp}$. We re-visit this scenario in Sec.~\ref{sec:numerical-simulation}.
% \end{remark}

%\ZBc{corrections to here 07/06/24 10:00}
% -------------------- Subsection 3.3 -----------------
\subsection{Recovery of EBSD maps}\label{subsec:compressive-ebsd-inpainting}
Our goal in this section is to recover a high quality estimate of the EBSD map, \ie $\hat{\bs z} \approx \bs x$, from an incomplete EBSD map $\bs z$ using inpainting. 

Inpainting is the recovery of an observation from fewer direct measurements. For image data, there are various inpainting methods based on, \eg interpolation ~\cite{chang2011new, wang2021electron}, deep learning ~\cite{elharrouss2020image,qin2021image}, and sparse coding ~\cite{zhou2009non,guillemot2013image} approaches. In this work, we consider beta-process factor analysis (BPFA): a joint dictionary learning and sparse coding algorithm from a sub-sampled set of measurements.

BPFA operates on a patch-wise basis. Given an observation $\bs{z}$ as in Eq.~\eqref{eq:indexing-model-compressive}, the image is broken down in to a set of $H_{\rm op} \times W_{\rm op}$ overlapping patches. The number of overlapping patches is thus $N_{\rm patch} = (H_{\rm p}-H_{\rm op}+1)(W_{\rm p}-W_{\rm op}+1)$. Each overlapping patch is then vectorised as $\bs z_{p} \in \bb R^{N_{\rm op}}$ with length $N_{\rm op} = H_{\rm op} W_{\rm op}$, forming a collection $\{\bs z_p\}_{p = 1}^{N_{\rm patch}}$ of all overlapping patches. Accordingly, the EBSD map $\bs x$, noise vector $\bs n$, and subsampling operator $\bs P_{\Omega}$ are partitioned and form the collections $\{\bs x_{p}\}_{p = 1}^{N_{\rm patch}}$, $\{\bs n_{p}\}_{p = 1}^{N_{\rm patch}}$, and $\{\bs P_{\Omega_{p}}\}_{p = 1}^{N_{\rm patch}}$, respectively. With those definitions, the acquisition model of each overlapping patch is given by
\begin{equation}\label{eq:indexing-model-compressive2}
    \bs z_{p} = \bs P_{\Omega_p}(\bs x_p + \bs n_p) \in \bb R^{N_{\rm op}}, {\rm ~for~} p \in \{1,\cdots,N_{\rm patch}\}.
\end{equation}
As required by BPFA, every patch of the EBSD map $\bs x_p \in \bb R^{N_{\rm op}}$ is assumed to have a sparse representation in a shared dictionary $\bs D \in \bb R^{N_{\rm op} \times K}$ of $K$ atoms, \ie for $p \in \{1,\cdots,N_{\rm patch}\}$,
\begin{equation}\label{eq:bpfa-sparsity}
    \bs x_p = \bs D \bs \alpha_p, {\rm ~with~} \|\bs \alpha_p \|_0 \le s,
\end{equation}
where $\bs \alpha_p\in \bb R^{K}$ denotes the sparse weight vector and $s$ is the sparsity level.

Other assumptions for BPFA include the following. \textit{(i)} Every dictionary atom $\bs d_k$, for $k \in \{1,\cdots,K\}$, is drawn from a zero-mean multivariate Gaussian distribution. \textit{(ii)} Both the components of the noise vectors $\bs n_p$ and the non-zero components of the sparse weight vectors $\bs \alpha_p$ are drawn \textit{i.i.d.} from zero-mean Gaussian distributions. \textit{(iii)} The sparsity prior on the weight vectors is promoted by the Beta-Bernoulli process~\cite{paisley2009nonparametric}. Those assumptions can be mathematically modelled in a hierarchical format as 
\begin{subequations}
\begin{align}
    \bs z_p &= \bs P_{\Omega_p} (\bs D \bs \alpha_p + \bs n_p), \!& \bs \alpha_p &= \bs u_p \odot \bs w_p \in \bb R^K,\label{eq:bpfa-1}\\
    \ts \bs D &= [\bs d_1^\top, \cdots, \bs d_{K}^\top]^{\top}, \!& \bs d_k & \sim  \cl N(0, \gamma_d^{-1} \bs I_{N_{\rm op}}),\label{eq:bpfa-2}\\
    \bs w_p & \sim  \cl N(0, \gamma_w^{-1} \bs I_{K}), \!& \bs n_p & \sim \cl N(0, \gamma_n^{-1} \bs I_{N_{\rm op}}),\label{eq:bpfa-3}\\
    \bs u_p &\sim \!\ts \prod_{k=1}^{K} {\rm Bernoulli}(\pi_k),\! & \pi_k &\sim\! {\rm Beta}(\ts \frac{a}{K}, \frac{b(K-1)}{K}),\label{eq:bpfa-4}
\end{align}
\end{subequations}
for every patch indexed by $p\in \{1,\cdots, N_{\rm patch}\}$ and every dictionary atom indexed $k \in \{1,\cdots,K\}$. In the equations above, the identity matrix of size $K\times K$ is denoted by $\bs I_K$, $\odot$ denotes the Hadamard product, and $a$ and $b$ are the parameters of the Beta process, $\bs u_p$ in is a binary vector controlling which dictionary atoms are used to represent $\bs x_p$. The probability of using a dictionary atom $\bs d_k$ for representing $\bs x_p$ is $\pi_k$. Moreover, $\gamma_d$, $\gamma_w$, and $\gamma_n$ are the precision parameters.

BPFA infers all the unknown parameters in the hierarchical model above. However, different methods can also be used for inference, such as, Gibbs sampling~\cite{zhou2009non} and variational inference~\cite{paisley2009nonparametric}. In this paper, we use a BPFA with Expectation Maximisation (EM)~\cite{dempster1977maximum} inference. See~\cite{sertoglu2015scalable} for more details on the BPFA-EM.

\section{Numerical results} \label{sec:numerical-simulation}
% Described here are the simulations performed and the results obtained from there. 
% \AMc{We should be more specific and avoid sentences that can be copied from here and pasted in other places.}

\sqp
\noindent\textbf{Fully sampled experimental data.} A fully sampled dataset -- \ie the collection of $\bs y_l$ for $l\in \{1,\cdots, N_{\rm p}\}$ in Eq.~\eqref{eq:forward-model} -- used for these simulations is a sample of a Ni-superalloy. Data was acquired on a Zeiss EVO15 SEM equipped with an Oxford Instruments Symmetry S3 detector at an accelerating voltage of 20 KeV, magnification of 1045x, exposure time of 0.9 ms, detector speed of 1057 Hz, and a total acquisition time of 3 min 21 s, with $(H_{\rm{p}},W_{\rm{p}}) = (512,416)$ probe positions. The EBSPs were binned during acquisition resulting $(H_{\rm{d}}, W_{\rm{d}}) = (156,128)$ detector pixels.

\sqp
\noindent\textbf{Reference EBSD maps.}
The reference band contrast and IPF Z maps, denoted by $\bs x$ in Eq.~\ref{eq:indexing-model-compressive} and shown in Figure~\ref{fig:mapex}, were generated using Hough-based indexing in AZtec from the fully sampled data, since this is a fast and readily available method. The indexing parameters used were optimised band detection with 11 bands and a Hough resolution of 40. 

No additional post-processing, \eg cleaning, was performed.

The above EBSD maps were selected due to their differing properties: the band contrast map is a greyscale image and the IPF Z map is a 3-D RGB image. There are also other EBSD maps which rely on a full EBSD data set, \eg kernel average misorientation maps where pixels are grouped together to assess misorientation ~\cite{EBSDinMatSci}

\sqp
\noindent\textbf{Quality criteria.}
Given a pair of noisy signal $\bs v$ and noiseless signal $\bs u$, the Signal-to-Noise-Ratio (SNR) in dB is defined as
\begin{equation}
    {\rm SNR} = 20\log_{10}\frac{\|\bs u\|}{\|\bs u-\bs v\|},
\end{equation}
where $\|u\| \coloneqq(\sum_{i=1}^{N}|u_i|^2)^{1/2}$ represents the $\ell_2$-norm of $\bs u$.
The robustness of Hough-based indexing is measured using a Hit Rate (HR) and normalised $\ell_2$-norm error. The hit rate is a measure of the percentage of EBSPs that have been successfully indexed.

\begin{equation}\label{eq:hit-rate-definition}
    {\rm HR} = 1- \frac{|\Omega_{\rm zsp}|}{N_{\rm p}}.
\end{equation}

The quality of the reconstructed maps are measured with respect to the reference maps using the Structural Similarity Index Measure (SSIM) ~\cite{wang2004image}.
% \AMc{Not sure atm whether we need to define SSIM.} \DNc{If we define SNR shouldn't be also define SSIM?}

\sqp
\noindent\textbf{Noise models.}
Two noise models were used in order to test the robustness of the EBSD indexing process. Gaussian noise and Poisson noise both affect microscopy data ~\cite{meiniel2018denoising, prasad2003sem}, with Gaussian noise most commonly encountered as detector readout noise ~\cite{meiniel2018denoising}, whilst Poisson noise is most encountered as counting noise, from the counting of electrons ~\cite{Joy2008}.

An independant Gaussian noise vector $\bs \eta \in \bb R^{N}$ of size $N$ was generated  following an independant and identically distributed (i.i.d.) Gaussian distribution, \ie $\eta_i \sim \cl N(0, \sigma_{\rm gsn})$ for $i \in \{1,\cdots, N\}$, with zero mean and standard deviation $\sigma_{\rm gsn}$. Therefore, given a noiseless signal $\bs y$ -- and considering one index of EBSPs in Eq.~\eqref{eq:forward-model} -- its Gaussian noisy version is generated as
\begin{equation}
    \bs y^{\rm gsn} = \bs y + \bs \eta.
\end{equation}
Given a desired SNR value in dB, we set $\sigma_{\rm gsn}$ as
\begin{equation}\label{eq:sigma_gaussian}
    \sigma_{\rm gsn} =\frac{\left \| \bs y \right \|}{\sqrt{N_{\rm{p}}}}\cdot 10^{-\frac{SNR}{20}}.
\end{equation}

Poisson noise is important in electron microscopy for counting direct electron detectors. The number of counts at a given detector pixel location is proportional to the scattering cross-section associated with that scattering angle ~\cite{echlin2013advanced,carter2016transmission}, which can be modelled by a Poisson distribution.

For this case let $\bs y^{\rm psn} \in \bb R^{N}$ be a noisy version of the noiseless signal $\bs y$ corrupted by Poisson noise. Therefore, $\bs y^{\rm psn}$ is a random vector of $N$ random variables with Poisson distribution, \ie $y_i^{\rm psn} = \cl P(y_i)$ for $i \in \{1,\cdots,N\}$. However, given a desired SNR value in dB, we can generate a noisy signal following
\begin{equation}
    y^{\rm psn}_i = \cl P\big(\sigma_{\rm psn} \frac{y_i}{\|\bs y\|_1}\big),
\end{equation}
where $\|\bs u\|_1 \coloneqq\sum_{i=1}^N|u_i|$ is the $\ell_1$-norm of vector $ \bs u \in \bb R^{N}$ and $\sigma_{\rm psn}$ controls the total absolute intensity of the noiseless signal. Assuming the vector $\bs y$ as one instance of EBSPs in Eq.~\eqref{eq:forward-model}, $\sigma_{\rm psn}$ models the total number of scattered electrons. A higher $\sigma_{\rm psn}$ corresponds to more electrons striking the detector. In our simulations, given a desired SNR we set $\sigma_{\rm psn}$ such that the resultant SNR is close to the desired SNR, following
% \AMc{Could you cross-validate this wrt to the values that manually were set?} \ZBc{I am going to do this I haven't yet}
\begin{equation}\label{eq:sigma-poisson}
    \sigma_{\rm psn} =\frac{\left \| \bs y \right \|_1^2}{\|\bs y\|^2}\cdot 10^{\frac{SNR}{10}}.
\end{equation}

\subsection{Robustness of indexing to noisy data}

\begin{figure*}[h]
	\centering
	\includegraphics[width=2\columnwidth]{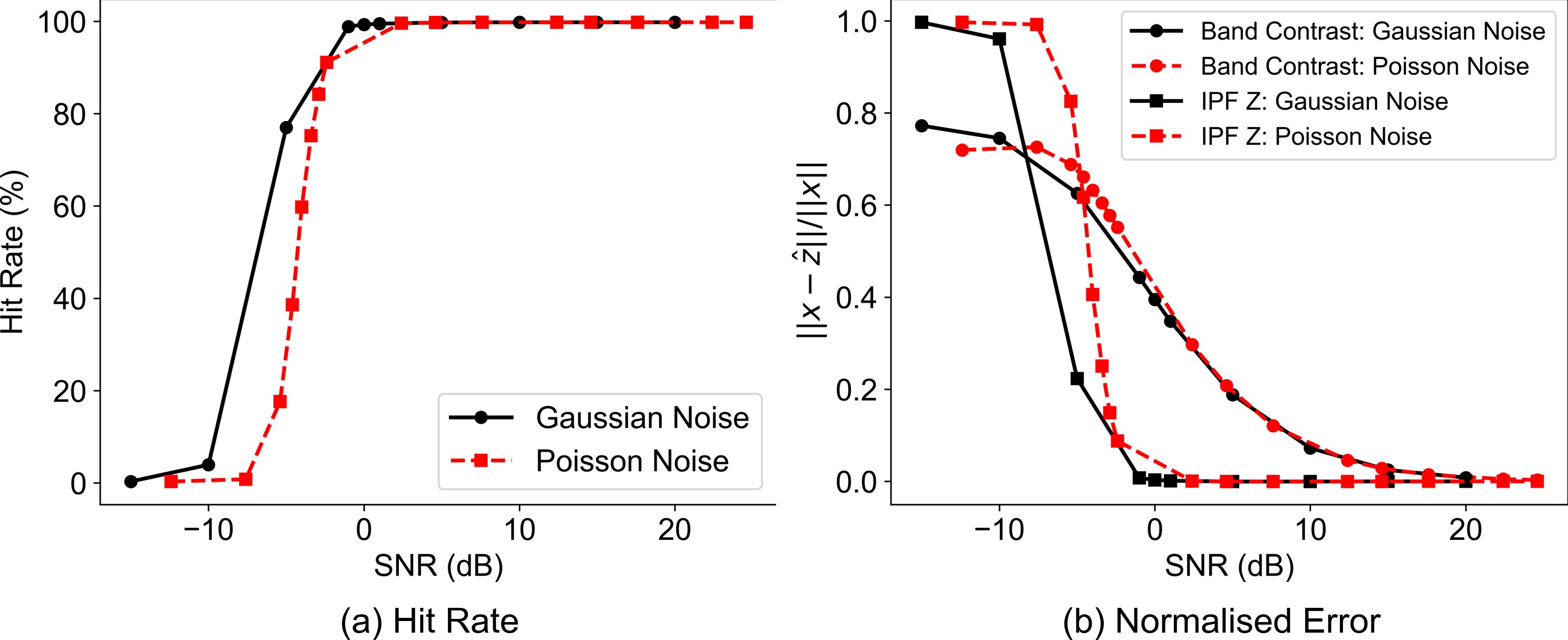}
	\caption{Effect of Gaussian and Poisson noise models on Hough-based indexing. (a) Hit rate;  
		(b) normalised error between reference and indexed EBSD maps.
		Due to the high amount of redundancy in EBSD data, the indexing process is robust to moderate noise levels. Datapoints from 5 and -5 dB are shown in Figure~\ref{fig:NoiseMaps}.} 
	\label{fig:hitrate}
\end{figure*}

In this section, Monte Carlo simulations are described with Gaussian and Poisson noise models for multiple independent realisations of noise with input SNR values in the range of $[-5, 5]$ dB. Noisy data sets were reprocessed using Hough-based indexing in the AZtec software. 
From this, the measured hit rate was recorded and plotted, as shown in Figure~\ref{fig:hitrate} (a).
This curve shows that for both noise models the hit rate remains above 99.5\% for input ${\rm SNR} >0$ dB. Note that for input ${\rm SNR} =0$ dB, signal and noise powers are equal. 
%As SNR drops below 0 dB the hit rate quickly decreases, reaching 0.3\% at ${\rm SNR} =-15$ dB for Gaussian noise and ${\rm SNR} =-12$ dB for Poisson noise. 

\begin{figure*}[h]
	\centering
	\includegraphics[width=2\columnwidth]{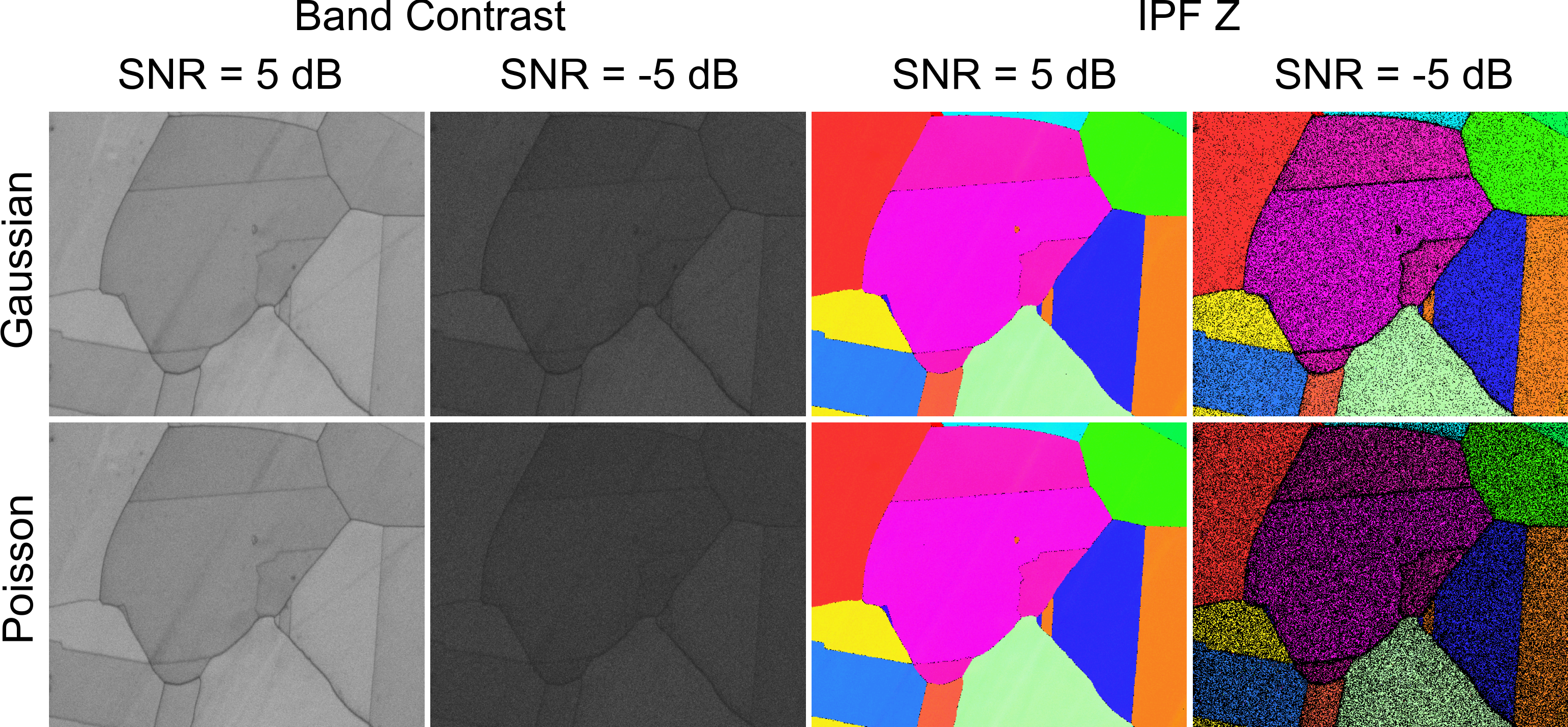}
	\caption{Band contrast and IPF Z maps produced through Hough-based indexing of noisy EBSD data corrupted by Gaussian and Poisson noise. The indexing quality of these maps are shown in Figure~\ref{fig:hitrate}.}
	\label{fig:NoiseMaps}
\end{figure*}

The normalised error between the reference and indexed EBSD maps is shown in Figure~\ref{fig:hitrate} (b). 

For the band contrast map, the normalised error slowly decreases for both noise models. In comparison, for the IPF Z map, the normalised error has a sharp drop when the SNR is between -8 dB and 0 dB. Those observations indicate the indexing processes involved for generating band contrast and IPF Z maps are robust to both Gaussian and Poisson noise models and yield high quality EBSD maps for moderate noise levels, \ie SNR values larger than 10 dB.

Examples of the EBSD maps associated with the SNR values of 5 dB and -5 dB are shown in Figure~\ref{fig:NoiseMaps}, where a lower hit rate is evident in the IPF Z map associated with the SNR = -5 dB, with more ZSPs being present across the map. For the same noise level, the band contrast map suffers from low contrast.
% --------------------------------

\subsection{Improving the robustness of IPF maps to strong noise}\label{subsec:numerical-result-zsp}

From the definition of ZSPs, and the observation in Figure~\ref{fig:hitrate}, our goal in this section is to correct the ZSPs in the IPF maps. We consider ZSPs as unsampled pixels; hence, the associated subsampled IPF map can be inpainted. That approach is referred to as ZSP correction. We emphasise that the subsampling of the IPF map here is not due to the probe subsampling. We will investigate that scenario momentarily.

\begin{figure*}
	\centering
	\includegraphics[width=2\columnwidth]{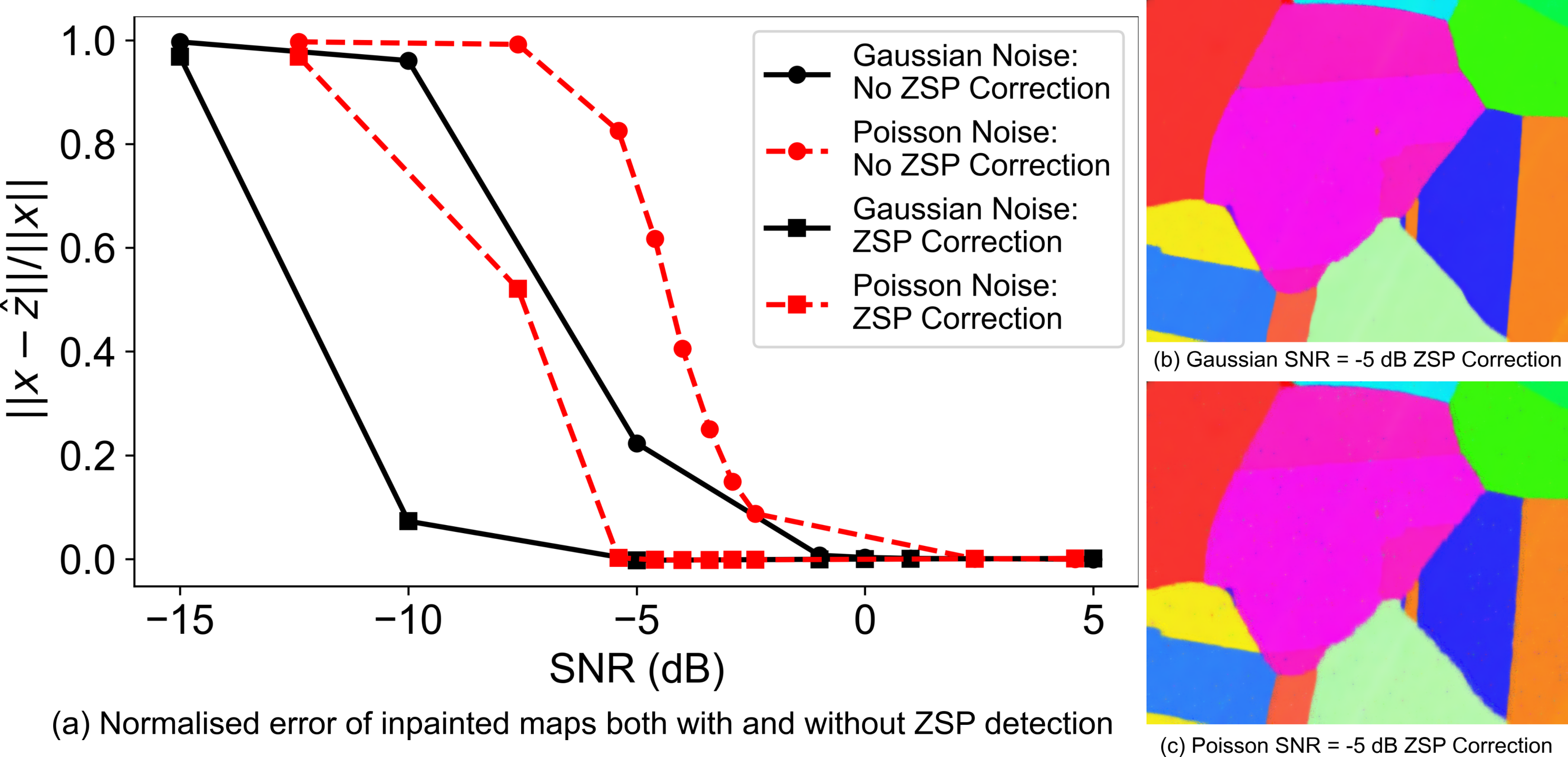}
	\caption{ (a) Normalised error of inpainted maps both with and without ZSP detection
		(b),(c) IPF Z maps with ZSP correction. Examples of these maps without ZSP correction can be found in Figure~\ref{fig:NoiseMaps}.}
	\label{fig:zspeffect}
\end{figure*}

Starting from Figure~\ref{fig:hitrate}, we limit the range of investigated SNR values to only a range where the hit rate is lower than 99\%. A comparison of normalised error of inpainted datasets with  ${\rm SNR}\in [-15,5]$~dB for both noise models, with and without ZSP correction is shown in Figure~\ref{fig:zspeffect} (a). We observe that ZSP correction significantly improves the indexing error, for example a 0.97 relative error improvement for Gaussian noise with ${\rm SNR} = -10$ dB and a 0.98 relative error improvement for Poisson noise with ${\rm SNR} = -7.6$.
Examples of the inpainted maps corresponding to SNR = -5 dB are shown in Figure~\ref{fig:zspeffect} (b) and (c). Those two maps should be compared with their counterparts in Figure~\ref{fig:NoiseMaps} with the same SNR values. It is evident that the maps from ZSP correction are fully inpainted -- the hit rate is now 100\% and show higher quality when compared to the reference map.

\subsection{Robustness of proposed method to subsampled noisy data}
In this part, we simulate a post-indexing reconstruction framework as shown in Figure~\ref{fig:ebsd_workflow} with both noisy and subsampled EBSD datasets generated, then indexed, and finally inpainted.
Following a UDS strategy of probe positions, subsampling rates $M_{\rm p}/N_{\rm p} \in \{1,5,10,15,20,25\} \%$ were investigated in conjunction with two input SNR values, \ie ${\rm SNR} \in \{-5,5\}$~dB for both Gaussian and Poisson noise models. For comparison, we have also included a noiseless case.

For computational reasons, band contrast maps were normalised to the range of [0, 255], a typical pixel range for an 8-bit image, prior to inpainting and subsequently re-normalised to their original range afterward.

Following the success of inpainting for correcting the ZSPs in the IPF map in Sec.~\ref{subsec:numerical-result-zsp}, those maps had an additional step performed before inpainting to identify the ZSPs, as explained previously. 
% \Rc{clarify that we do not include zero solution pixels in the calculation of subsampling}
Here we do not include the zero solution pixels in the calculation of subsampling rate, therefore, the effective subsampling rate -- \ie that used by the inpainting algorithm -- corresponding to IPF map simulations is lower than the desired subsampling rate set according to $M_{\rm p}/N_{\rm p}$ ratios mentioned above. More precisely, the effective subsampling rate will be in the range of $[\frac{|\Omega_{\rm zsp}| + M_{\rm p}}{N_{\rm p}}, \frac{M_{\rm p}}{N_{\rm p}}]$. A previous work~\cite{broad2023subsampling} did not include this step, resulting in untreated ZSPs, which in turn limited the reconstruction quality.

Effective inpainting parameters for the BPFA algorithm were selected based on a parameter optimisation routine consisting of a series of full factorial design of experiment trials. 
The common parameters for all simulations were as follows: dictionary atoms $K = 25$, sparsity limit $s = 4$, batch size of 1024, and only one epoch, \ie one pass over the full dataset.

It has been previously demonstrated that patch shape has the greatest effect on the outcome of a BPFA reconstruction ~\cite{nicholls2022compressive}. Therefore, the patch shapes were selected individually for each subsampling rate and each EBSD map type. These are shown in Table \ref{tab:table}. 

\begin{table}
	\centering
	\caption{Patch shape selected at each sampling rate for inpainting.}
	\begin{tabular}{ccc}
		&  \multicolumn{2}{c}{\textbf{Patch Shape }$[H_{\rm op},W_{\rm op}]$} \\
		\textbf{Sampling Rate} (\%)&  \textbf{Band Contrast} & \textbf{IPF Z}\\
		% \hline
		25&  [6, 6]& [9, 9]\\
		20&  [8, 8]& [11, 11]\\
		15&  [8, 8]& [11, 11]\\
		10&  [10, 10]& [13, 13]\\
		5&  [16, 16]& [14, 14]\\
		1&  [27, 27]& [23, 23]\\
		% \hline
	\end{tabular}
	\label{tab:table}
\end{table}
 
\begin{figure*}[t]
	\centering
	\includegraphics[width=2\columnwidth]{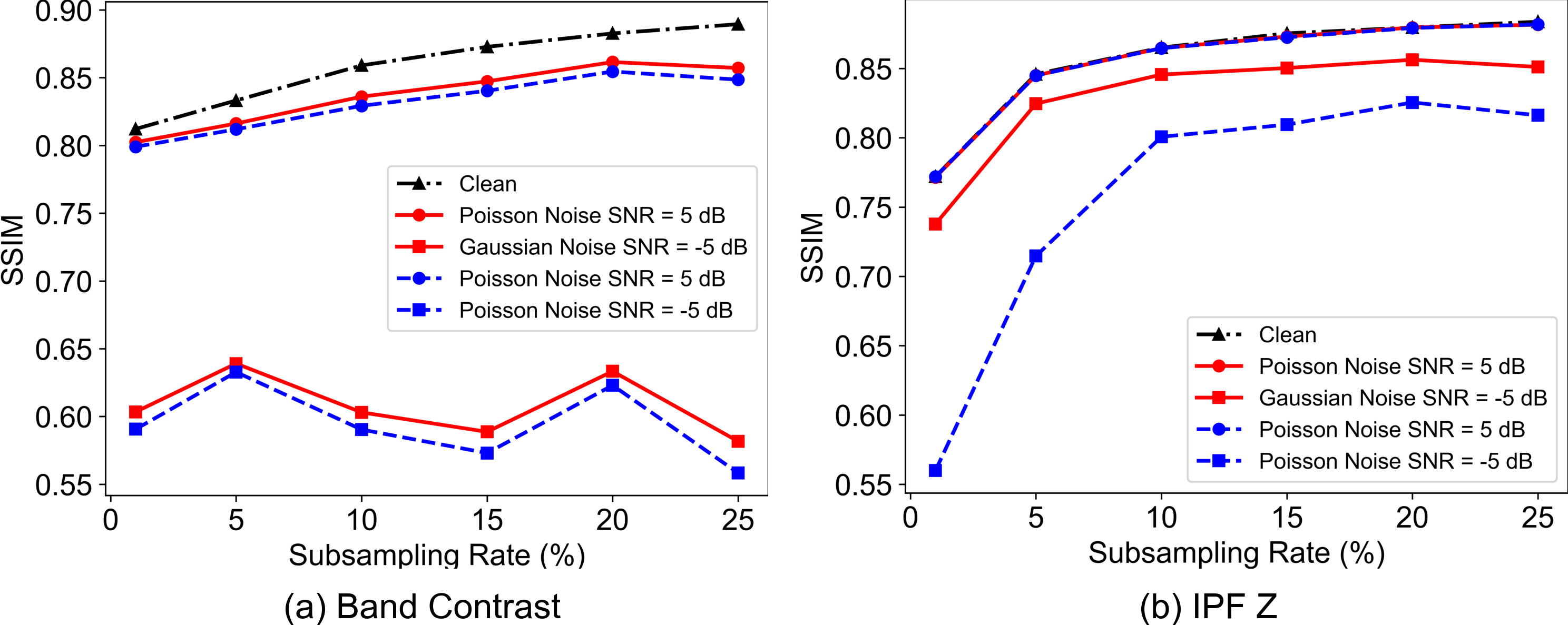}
	\caption{Image quality of reconstructed subsampled EBSD maps showing the effects of noise on the quality of the images reconstructed.
		(a) band contrast; (b) IPF Z maps.}
	\label{fig:bc_ipf_results}
\end{figure*}
 
The inpainting results are shown in Figure~\ref{fig:bc_ipf_results} for the band contrast maps.  the inpainting results show a steady increase from 1\% to 10\%, which then plateaus as the subsampling rate increases further. The same trend is found for the noisy datasets when ${\rm SNR} = 5$ dB, for both noise models. The plateau suggests that it should be possible to subsample EBSD datasets down to 10\%, with band contrast maps reconstructed with a quality comparable to that achieved from a full EBSD dataset.
However, for input ${\rm SNR} = -5$ dB, there is a significant detrimental impact on the inpainting results, which was expected from the data in Figure~\ref{fig:hitrate}: indexing an EBSD dataset with such strong noise gives a normalised error of approximately 0.7.

The IPF Z results are shown in Figure~\ref{fig:bc_ipf_results}.
An immediate observation is that inpainting IPF maps is more robust to noisy and subsampled data compared to that of band contrast maps for which the plateau starts at 5\% subsampling rate. The reason for this is twofold. Firstly, an IPF map is a three-colour map with more diversity; hence, helping the inpainting algorithm improve the reconstruction quality. Secondly, ZSPs are considered as unsampled probe positions. Therefore, those pixels are corrected during inpainting and the SSIM values for the clean dataset and the dataset with ${\rm SNR} = 5$ dB are almost identical. For these maps a lower SSIM than the band contrast maps is obtained at a very low subsampling ratio of 1\%.  The proposed method shows reasonable robustness to a very strong Gaussian noise (${\rm SNR} = -5$ dB), with its SSIM curve slightly shifted down. However, a dataset corrupted by a very strong Poisson noise (${\rm SNR} = -5$ dB) gives lower SSIM values, which is likely due to its lower hit rate (39\% compared to 77\% for the Gaussian noise) as shown in Figure~\ref{fig:hitrate}.
Overall, those results suggest that subsampling only 10\% of probe positions and inpainting the missing data in the band contrast and IPF maps yields high quality maps, even if the data is corrupted by a moderate Gaussian or Poisson noise, \ie with ${\rm SNR} = 5$ dB.

\begin{figure*}
	\centering
	\includegraphics[width=2\columnwidth]{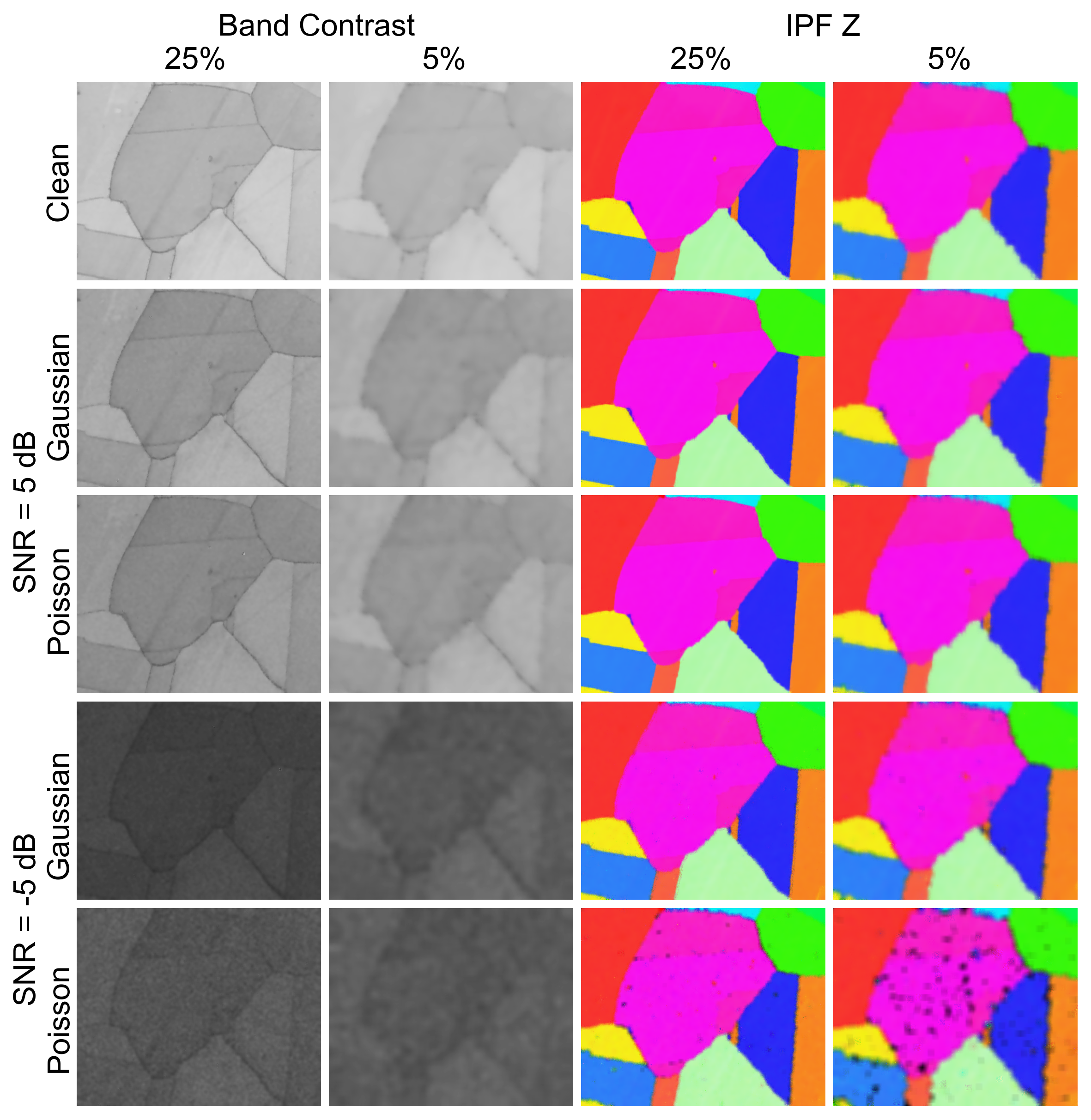}
	\caption{Examples of inpainted band contrast and IPF Z maps. These datapoints are taken from Figure~\ref{fig:bc_ipf_results}.}
	\label{fig:ReconMaps}
\end{figure*}

Examples of the inpainted clean and noisy band contrast and IPF Z maps are shown in Figure~\ref{fig:ReconMaps}. The effect of the lower subsampling rate is evident with more blurring being present at 5\% than at 25\%. The effect of noise is also seen in the band contrast maps, with a darker map being produced by noisier datasets. 

\section{Conclusion and Discussion}

We have presented a proof of concept for compressive EBSD imaging. We show the efficacy of subsampling and inpainting applied to both clean and noisy EBSD datasets.

Using BPFA as an inpainting method, clean datasets can be subsampled to 10\%, with the potential to reduce this to 5\% if only IPF Z maps will be are required. Datasets with moderate noise levels have a similar robustness to subsampling, with the band contrast showing robustness to 10\% subsampling and IPF Z to 5\% subsampling. Overall, we have shown that ZSP correction via inpainting enables higher quality reconstructions of IPF maps when the EBSD data set is corrupted by strong noise.  

Compared to the proposed method based on post-indexing reconstruction, an alternative mechanism would be based on pre-indexing reconstruction involving inpainting the EBSPs, forming a complete 4-D EBSD dataset for indexing. This alternative mechanism of reconstruction is expected to be a more robust method of inpainting EBSD datasets, allowing fast and low dose acquisition suitable for more beam sensitive samples. Furthermore, leveraging the high amount of diversity in the 4-D EBSD data set during the inpainting process would enable noisier datasets, equivalent to lower electron dose, to be collected.

These results show how subsampling can be applied to EBSD datasets. Whilst these results are positive, further work to ensure robustness is necessary. These include but are not limited to experimentally noisy datasets; small grains and defects/deformations.

Although this method is robust to the noise models shown here, its equivalency to experimental noise will be investigated. Experimental noise tends to result in greater noise levels around the grain boundaries reflected by more ZSPs. This should not prevent the inpainting algorithm from being effective, although it could reduce the resolution at grain boundaries. As shown earlier, there is blur in the reconstructions which may increase where less information is available. 
Alternatively, experimentally noisy datasets can result in mis-indexed pixels. Although it's possible that the inpainting algorithm could remove low levels of mis-indexing, where many mis-indexed pixels are present the method will be less robust, since the learned information will be inaccurate. It is anticipated that for these types of datasets, reconstructing the full dataset through pre-indexing reconstruction will be more robust, since the EBSPs will be subject to denoising.

Although the dataset shown here contains mostly larger grains, the small grains reconstructed in the EBSD map show that the algorithm has the sensitivty to small grains. Where small grains are present the most important aspect of this method is the selection of sampling rate, since any small grains that are not sampled will not be reconstructed. This method will be investigated on further samples, notably geological samples containing multiple phases, as well as an alternative inpainting method such as pre-indexing reconstruction to determine the best methods for sensitivity and speed.

Samples containing deformations or defects will also be investigated.

High angular resolution EBSD is one specific technique that could benefit from proposed framework. High angular resolution EBSD requires high resolution EBSPs in order to measure acute angles between expected and experimental bands in EBSPs ~\cite{wilkinson2012strains}. These high resolution EBSPs therefore have long acquisition times. By subsampling probe positions significant time could be saved during the acquisition.

Similarly, 3-D EBSD is another technique with lengthy acquisition times. In 3-D EBSD layers of a thick sample are cut, with a full EBSD map being acquired at each layer ~\cite{wilkinson2012strains}. That technique is limited by both cutting time and acquisition time, which could be significantly decreased through an application of probe position or layer subsampling. In that context, targeted sampling, which has been previously demonstrated in ~\cite{nicholls2023targeted}, would be a potential method for locating regions of interest such as grain boundaries. 

\section{Acknowledgements}
Louise Hughes and Robert Masters from Oxford Instruments NanoAnalysis are thanked for their input to this work.

\vfill\pagebreak    

\bibliographystyle{IEEEtran}
\bibliography{references}

     % Postscript figures can be included with multiple figure blocks

%\onecolumn

% \newpage

\end{document}

%% file: packages.tex
\usepackage{graphicx}   % Required for inserting images
\usepackage[sort&compress, numbers]{natbib} % Used for the handling of references and bibliography.
\usepackage{authblk}    % Used for handling authors and affiliations.
\usepackage{hyperref}   % Used for creating hyperlinks within document.
\usepackage{xcolor}     % Used for creating coloured text.

\usepackage{multirow}
\usepackage{lscape}

\usepackage{overpic}

\usepackage{amssymb,amsthm,amsmath,array}
\usepackage[mode=buildnew]{standalone}
\usepackage{graphicx}
\graphicspath{{Figures/}}
\usepackage{xspace}
\usepackage[sort&compress, numbers]{natbib}
\usepackage{stmaryrd}
\usepackage{xcolor}
\usepackage{mathtools}
\usepackage{float}
\usepackage{textcomp}
\usepackage{caption}
\usepackage{subcaption}
\usepackage{gensymb}

\usepackage{amssymb,amsthm,amsmath,array}
\usepackage{graphicx}
\usepackage{mathtools}
\usepackage{float}
\usepackage{textcomp}
\usepackage{comment}
\usepackage{soul} % Added by AM
\usepackage{todonotes} % Added by AM
\usepackage{standalone}
\graphicspath{{Figures/}}

\usepackage{tikz}
\usetikzlibrary{external}
\usetikzlibrary{arrows}
\usepackage{pgfplots}
\pgfplotsset{compat=1.7}

\newcommand{\bs}{\boldsymbol}
\newcommand{\bb}{\mathbb}
\newcommand{\cl}{\mathcal}

\newcommand{\ts}{\textstyle}
\newcommand{\ie}{\emph{i.e.},\xspace}
\newcommand{\eg}{\emph{e.g.},\xspace}
\newcommand{\etal}{\emph{et al.}}

\newcommand{\sqp}{\vspace{2mm}}

%% file: new_main.bbl
\begin{thebibliography}{10}
\providecommand{\url}[1]{#1}
\csname url@samestyle\endcsname
\providecommand{\newblock}{\relax}
\providecommand{\bibinfo}[2]{#2}
\providecommand{\BIBentrySTDinterwordspacing}{\spaceskip=0pt\relax}
\providecommand{\BIBentryALTinterwordstretchfactor}{4}
\providecommand{\BIBentryALTinterwordspacing}{\spaceskip=\fontdimen2\font plus
\BIBentryALTinterwordstretchfactor\fontdimen3\font minus
  \fontdimen4\font\relax}
\providecommand{\BIBforeignlanguage}[2]{{%
\expandafter\ifx\csname l@#1\endcsname\relax
\typeout{** WARNING: IEEEtran.bst: No hyphenation pattern has been}%
\typeout{** loaded for the language `#1'. Using the pattern for}%
\typeout{** the default language instead.}%
\else
\language=\csname l@#1\endcsname
\fi
#2}}
\providecommand{\BIBdecl}{\relax}
\BIBdecl

\bibitem{EBSDinMatSci}
A.~J. Schwartz, M.~Kumar, B.~L. Adams, and D.~P. Field, \emph{Electron
  backscatter diffraction in materials science}, 2nd~ed.\hskip 1em plus 0.5em
  minus 0.4em\relax New York: Springer, 2009.

\bibitem{hough1962method}
P.~V. Hough, ``Method and means for recognizing complex patterns,'' Dec.~18
  1962, uS Patent 3,069,654.

\bibitem{lassen1994automated}
N.~C.~K. Lassen, ``Automated determination of crystal orientations from
  electron backscattering patterns,'' Ph.D. dissertation, The Technical
  University of Denmark, 1994.

\bibitem{chen2015dictionary}
Y.~H. Chen, S.~U. Park, D.~Wei, G.~Newstadt, M.~A. Jackson, J.~P. Simmons,
  M.~De~Graef, and A.~O. Hero, ``A dictionary approach to electron backscatter
  diffraction indexing,'' \emph{Microscopy and Microanalysis}, vol.~21, no.~3,
  pp. 739--752, 2015.

\bibitem{lenthe2019spherical}
W.~Lenthe, S.~Singh, and M.~De~Graef, ``A spherical harmonic transform approach
  to the indexing of electron back-scattered diffraction patterns,''
  \emph{Ultramicroscopy}, vol. 207, p. 112841, 2019.

\bibitem{rowenhorst2024fast}
D.~J. Rowenhorst, P.~G. Callahan, and H.~W. {\AA}nes, ``Fast radon transforms
  for high-precision ebsd orientation determination using pyebsdindex,''
  \emph{Journal of Applied Crystallography}, vol.~57, no.~1, 2024.

\bibitem{wright2006ebsd}
S.~I. Wright and M.~M. Nowell, ``Ebsd image quality mapping,'' \emph{Microscopy
  and microanalysis}, vol.~12, no.~1, pp. 72--84, 2006.

\bibitem{SEMGoldstein}
J.~I. Goldstein, D.~E. Newbury, J.~R. Michael, N.~W. Ritchie, J.~H.~J. Scott,
  and D.~C. Joy, \emph{Scanning electron microscopy and X-ray
  microanalysis}.\hskip 1em plus 0.5em minus 0.4em\relax Springer, 2017.

\bibitem{EDAXCMOS}
M.~M. Nowell, ``Cutting-edge ebsd detector technology,'' \emph{Wiley Analytical
  Science}, 2022.

\bibitem{sertoglu2015scalable}
S.~Sertoglu and J.~Paisley, ``Scalable bayesian nonparametric dictionary
  learning,'' in \emph{2015 23rd European Signal Processing Conference
  (EUSIPCO)}, 2015, pp. 2771--2775.

\bibitem{paisley2009nonparametric}
J.~Paisley and L.~Carin, ``Nonparametric factor analysis with beta process
  priors.'' \emph{Proceedings of the 26th annual international conference on
  machine learning}, pp. 777--784, 2009.

\bibitem{candes2006robust}
E.~J. Candès, J.~Romberg, and T.~Tao, ``Robust uncertainty principles: Exact
  signal reconstruction from highly incomplete frequency information.''
  \emph{IEEE Transactions on information theory}, vol. 52(2), pp. 489--509,
  2006.

\bibitem{donoho2006compressed}
D.~L. Donoho, ``Compressed sensing,'' \emph{IEEE Transactions on information
  theory}, vol. 52(4), pp. 1289--1306, 2006.

\bibitem{mallat1999wavelet}
S.~Mallat, \emph{A wavelet tour of signal processing}.\hskip 1em plus 0.5em
  minus 0.4em\relax Elsevier, 1999.

\bibitem{saghi2015reduced}
Z.~Saghi, M.~Benning, R.~Leary, M.~Macias-Montero, A.~Borras, and P.~A.
  Midgley, ``Reduced-dose and high-speed acquisition strategies for
  multi-dimensional electron microscopy,'' \emph{Advanced Structural and
  Chemical Imaging}, vol.~1, no.~1, pp. 1--10, 2015.

\bibitem{kovarik2016implementing}
L.~Kovarik, A.~Stevens, A.~Liyu, and N.~D. Browning, ``Implementing an accurate
  and rapid sparse sampling approach for low-dose atomic resolution stem
  imaging,'' \emph{Applied Physics Letters}, vol. 109, no.~16, 2016.

\bibitem{mucke2016practical}
D.~Mucke-Herzberg, P.~Abellan, M.~C. Sarahan, I.~S. Godfrey, Z.~Saghi, R.~K.
  Leary, A.~Stevens, J.~Ma, G.~Kutyniok, F.~Azough \emph{et~al.}, ``Practical
  implementation of compressive sensing for high resolution stem,''
  \emph{Microscopy and Microanalysis}, vol.~22, no.~S3, pp. 558--559, 2016.

\bibitem{nicholls2022compressive}
D.~Nicholls, A.~W. Robinson, J.~Wells, A.~Moshtaghpour, M.~Bahri, A.~I.
  Kirkland, and N.~D. Browning, ``Compressive scanning transmission electron
  microscopy,'' \emph{ICASSP 2022-2022 IEEE International Conference on
  Acoustics, Speech and Signal Processing (ICASSP)}, pp. 1586--1590, 2022.

\bibitem{nicholls2022sub}
D.~Nicholls, J.~Wells, A.~Stevens, Y.~Zheng, J.~Castagna, and N.~D. Browning,
  ``Sub-sampled imaging for stem: Maximising image speed, resolution and
  precision through reconstruction parameter refinement.''
  \emph{Ultramicroscopy}, vol. 233, p. 113451, 2022.

\bibitem{browning2023advantages}
N.~D. Browning, J.~Castagna, A.~I. Kirkland, A.~Moshtaghpour, D.~Nicholls,
  A.~W. Robinson, J.~Wells, and Y.~Zheng., ``The advantages of sub-sampling and
  inpainting for scanning transmission electron microscopy,'' \emph{Applied
  Physics Letters}, vol. 122(5), p. 050501, 2023.

\bibitem{robinson2023silico}
A.~W. Robinson, A.~Moshtaghpour, J.~Wells, D.~Nicholls, Z.~Broad, A.~I.
  Kirkland, B.~L. Mehdi, and N.~D. Browning, ``{In silico Ptychography of
  Lithium-ion Cathode Materials from Subsampled 4-D STEM Data},'' \emph{arXiv
  preprint arXiv:2307.06138}, 2023.

\bibitem{robinson2023simultaneous}
A.~W. Robinson, A.~Moshtaghpour, J.~Wells, D.~Nicholls, M.~Chi, I.~MacLaren,
  A.~I. Kirkland, and N.~D. Browning, ``Simultaneous high-speed and low-dose
  4-d stem using compressive sensing techniques,'' 2023.

\bibitem{moshtaghpour2023exploring}
A.~Moshtaghpour, A.~Velazco-Torrejon, A.~W. Robinson, A.~I. Kirkland, and N.~D.
  Browning, ``Exploring low-dose and fast electron ptychography using l0
  regularisation of extended ptychographical iterative engine,'' 2023.

\bibitem{robinson2023developing}
A.~Robinson, ``Developing computational imaging methods for quantitative
  multi-dimensional electron microscopy,'' Ph.D. dissertation, University of
  Liverpool, 2023.

\bibitem{robinson2023towards}
A.~W. Robinson, J.~Wells, D.~Nicholls, A.~Moshtaghpour, M.~Chi, A.~I. Kirkland,
  and N.~D. Browning, ``{Towards real-time STEM simulations through targeted
  subsampling strategies},'' \emph{Journal of microscopy}, vol. 290, no.~1, pp.
  53--66, 2023.

\bibitem{robinson2022sim}
A.~W. Robinson, D.~Nicholls, J.~Wells, A.~Moshtaghpour, A.~Kirkland, and N.~D.
  Browning, ``{SIM-STEM Lab: Incorporating compressed sensing theory for fast
  STEM simulation},'' \emph{Ultramicroscopy}, vol. 242, p. 113625, 2022.

\bibitem{nicholls2023targeted}
D.~Nicholls, J.~Wells, A.~W. Robinson, A.~Moshtaghpour, M.~Kobylynska, R.~A.
  Fleck, A.~I. Kirkland, and N.~D. Browning, ``A targeted sampling strategy for
  compressive cryo focused ion beam scanning electron microscopy,'' \emph{arXiv
  preprint arXiv:2211.03494}, 2022.

\bibitem{nicholls2023potential}
D.~Nicholls, M.~Kobylysnka, J.~Wells, Z.~Broad, D.~McGrouther, R.~A. Fleck, and
  N.~D. Browning, ``The potential of subsampling and inpainting for fast
  low-dose cryo fib-sem imaging and tomography,'' \emph{arXiv preprint
  arXiv:2309.09617}, 2023.

\bibitem{yang1999adaptive}
W.~Yang, C.-T. Wu, B.~Adams, and M.~De~Graef, ``Adaptive orientation imaging
  microscopy,'' \emph{Microscopy and microanalysis}, vol.~5, no.~S2, pp.
  246--247, 1999.

\bibitem{tong2019rapid}
V.~S. Tong, A.~J. Knowles, D.~Dye, and T.~B. Britton, ``Rapid electron
  backscatter diffraction mapping: Painting by numbers,'' \emph{Materials
  Characterization}, vol. 147, pp. 271--279, 2019.

\bibitem{long2024high}
T.~J. Long, W.~Holbrook, T.~C. Hufnagel, and T.~Mueller, ``High-throughput
  determination of grain size distributions by ebsd with low-discrepancy
  sampling,'' \emph{Journal of Microscopy}, vol. 293, no.~1, pp. 20--37, 2024.

\bibitem{wang2021electron}
F.~Wang, M.~P. Echlin, A.~A. Taylor, J.~Shin, B.~Bammes, B.~D. Levin,
  M.~De~Graef, T.~M. Pollock, and D.~S. Gianola, ``Electron backscattered
  diffraction using a new monolithic direct detector: High resolution and fast
  acquisition,'' \emph{Ultramicroscopy}, vol. 220, p. 113160, 2021.

\bibitem{wright2006random}
S.~Wright, ``Random thoughts on non-random misorientation distributions,''
  \emph{Materials Science and Technology}, vol.~22, no.~11, pp. 1287--1296,
  2006.

\bibitem{brewick2019nlpar}
P.~T. Brewick, S.~I. Wright, and D.~J. Rowenhorst, ``Nlpar: Non-local smoothing
  for enhanced ebsd pattern indexing,'' \emph{Ultramicroscopy}, vol. 200, pp.
  50--61, 2019.

\bibitem{brewer2008multivariate}
L.~N. Brewer, P.~G. Kotula, and J.~R. Michael, ``Multivariate statistical
  approach to electron backscattered diffraction,'' \emph{Ultramicroscopy},
  vol. 108, no.~6, pp. 567--578, 2008.

\bibitem{wilkinson2019applications}
A.~J. Wilkinson, D.~M. Collins, Y.~Zayachuk, R.~Korla, and A.~Vilalta-Clemente,
  ``Applications of multivariate statistical methods and simulation libraries
  to analysis of electron backscatter diffraction and transmission kikuchi
  diffraction datasets,'' \emph{Ultramicroscopy}, vol. 196, pp. 88--98, 2019.

\bibitem{brewer2010risks}
L.~Brewer and J.~Michael, ``Risks of “cleaning” electron backscatter
  diffraction data,'' \emph{Microscopy Today}, vol.~18, no.~2, p. 10–15,
  2010.

\bibitem{singh2017application}
S.~Singh, F.~Ram, and M.~De~Graef, ``Application of forward models to crystal
  orientation refinement,'' \emph{Journal of Applied Crystallography}, vol.~50,
  no.~6, pp. 1664--1676, 2017.

\bibitem{wright1992automatic}
S.~I. Wright and B.~L. Adams, ``Automatic analysis of electron backscatter
  diffraction patterns,'' \emph{Metallurgical Transactions A}, vol.~23, pp.
  759--767, 1992.

\bibitem{ram2017error}
F.~Ram, S.~Wright, S.~Singh, and M.~De~Graef, ``Error analysis of the crystal
  orientations obtained by the dictionary approach to ebsd indexing,''
  \emph{Ultramicroscopy}, vol. 181, pp. 17--26, 2017.

\bibitem{godaliyadda2017supervised}
G.~M. D.~P. Godaliyadda, ``A supervised learning approach for dynamic sampling
  (slads),'' Ph.D. dissertation, Purdue University, 2017.

\bibitem{kandel2023demonstration}
S.~Kandel, T.~Zhou, A.~V. Babu, Z.~Di, X.~Li, X.~Ma, M.~Holt, A.~Miceli,
  C.~Phatak, and M.~J. Cherukara, ``Demonstration of an ai-driven workflow for
  autonomous high-resolution scanning microscopy,'' \emph{Nature
  Communications}, vol.~14, no.~1, p. 5501, 2023.

\bibitem{chang2011new}
L.~Chang and Y.~Chongxiu, ``New interpolation algorithm for image inpainting,''
  \emph{Physics Procedia}, vol.~22, pp. 107--111, 2011.

\bibitem{elharrouss2020image}
O.~Elharrouss, N.~Almaadeed, S.~Al-Maadeed, and Y.~Akbari, ``Image inpainting:
  A review,'' \emph{Neural Processing Letters}, vol.~51, pp. 2007--2028, 2020.

\bibitem{qin2021image}
Z.~Qin, Q.~Zeng, Y.~Zong, and F.~Xu, ``Image inpainting based on deep learning:
  A review,'' \emph{Displays}, vol.~69, p. 102028, 2021.

\bibitem{zhou2009non}
M.~Zhou, H.~Chen, L.~Ren, G.~Sapiro, L.~Carin, and J.~Paisley, ``Non-parametric
  bayesian dictionary learning for sparse image representations,''
  \emph{Advances in neural information processing systems}, vol.~22, 2009.

\bibitem{guillemot2013image}
C.~Guillemot and O.~Le~Meur, ``Image inpainting: Overview and recent
  advances,'' \emph{IEEE signal processing magazine}, vol.~31, no.~1, pp.
  127--144, 2013.

\bibitem{dempster1977maximum}
A.~P. Dempster, N.~M. Laird, and D.~B. Rubin, ``Maximum likelihood from
  incomplete data via the em algorithm,'' \emph{Journal of the Royal
  Statistical Society: Series B (Methodological)}, vol.~39, no.~1, pp. 1--22,
  1977.

\bibitem{wang2004image}
Z.~Wang, A.~C. Bovik, H.~R. Sheikh, and E.~P. Simoncelli, ``Image quality
  assessment: from error visibility to structural similarity,'' \emph{IEEE
  transactions on image processing}, vol.~13, no.~4, pp. 600--612, 2004.

\bibitem{meiniel2018denoising}
W.~Meiniel, J.-C. Olivo-Marin, and E.~D. Angelini, ``Denoising of microscopy
  images: a review of the state-of-the-art, and a new sparsity-based method,''
  \emph{IEEE Transactions on Image Processing}, vol.~27, no.~8, pp. 3842--3856,
  2018.

\bibitem{prasad2003sem}
M.~S. Prasad and D.~C. Joy, ``Is sem noise gaussian?'' \emph{Microscopy and
  Microanalysis}, vol.~9, no. S02, pp. 982--983, 2003.

\bibitem{Joy2008}
D.~C. Joy, \emph{Biological Low-Voltage Scanning Electron Microscopy}.\hskip
  1em plus 0.5em minus 0.4em\relax Springer New York, 2008, ch. Noise and Its
  Effects on the Low-Voltage SEM.

\bibitem{echlin2013advanced}
P.~Echlin, C.~Fiori, J.~Goldstein, D.~C. Joy, and D.~E. Newbury, \emph{Advanced
  scanning electron microscopy and X-ray microanalysis}.\hskip 1em plus 0.5em
  minus 0.4em\relax Springer Science \& Business Media, 2013.

\bibitem{carter2016transmission}
C.~B. Carter and D.~B. Williams, \emph{Transmission electron microscopy:
  Diffraction, imaging, and spectrometry}.\hskip 1em plus 0.5em minus
  0.4em\relax Springer, 2016.

\bibitem{broad2023subsampling}
Z.~Broad, D.~Nicholls, J.~Wells, A.~W. Robinson, A.~Moshtaghpour, R.~Masters,
  L.~Hughes, and N.~D. Browning, ``Subsampling methods for fast electron
  backscattered diffraction analysis,'' \emph{arXiv preprint arXiv:2307.08480},
  2023.

\bibitem{wilkinson2012strains}
A.~J. Wilkinson and T.~B. Britton, ``Strains, planes, and ebsd in materials
  science,'' \emph{Materials today}, vol.~15, no.~9, pp. 366--376, 2012.

\end{thebibliography}
